\documentclass[onecolumn,preprintnumbers,amsmath,amssymb]{revtex4}

\usepackage{graphicx}
\usepackage[latin1]{inputenc}
\usepackage{dcolumn}
\usepackage{bm}
\usepackage{epsfig}
\usepackage{color}

\begin{document}


\title{Visibility graphs of random scalar fields and spatial data}

\author{Lucas Lacasa and Jacopo Iacovacci}

\affiliation{School of Mathematical Sciences, Queen Mary University of London, Mile End Road, London E14NS (UK)}%

\date{\today}

\begin{abstract}
The family of visibility algorithms were recently introduced (Lacasa et al, PNAS 105 (2008)) as mappings between time series and graphs. Here we extend this method to characterize spatially extended data structures by mapping scalar fields of arbitrary dimension into graphs. After introducing several possible extensions, we provide analytical results on some topological properties of these graphs
associated to some types of real-valued matrices, which can be understood as the high and low disorder limits of real-valued scalar fields. In particular, we find a closed expression for the degree distribution of these graphs associated to uncorrelated random fields of generic dimension, extending a well known result in one-dimensional time series. As this result holds independently of the field's marginal distribution, we show that it directly yields a statistical randomness test, applicable in any dimension. We showcase its usefulness by discriminating spatial snapshots of two-dimensional white noise from snapshots of a two-dimensional lattice of diffusively coupled chaotic maps, a system that generates high dimensional spatio-temporal chaos.
We finally discuss the range of potential applications of this combinatorial framework, which include image processing in engineering, the description of surface growth in material science, soft matter or medicine and the characterization of potential energy surfaces in chemistry, disordered systems and high energy physics. An illustration on the applicability of this method for the classification of the different stages involved in carcinogenesis is briefly discussed.
\end{abstract}

\maketitle

\section{Introduction}
Visibility and Horizontal Visibility Graphs are a family of mappings between ordered sequences and graphs \cite{PNAS, PRE}. Consider an ordered sequence $\{{\bf x}(t)\}_{t=1}^N$, where ${\bf x}(t)\in \mathbb{R}^m, \ m\geq 1$. For $m=1$ a typical case of such sequence are time series describing the activity of some system, whereas for $m>1$ we consider multivariate time series or in general high-dimensional dynamical systems. In every case, a univariate time series of $N$ data is mapped into a graph of $N$ nodes such that two nodes are linked in the graph if a particular visibility criterion holds in the sequence (the multivariate setting has been explored recently \cite{multivariate}). This mapping enables the possibility of performing graph-theoretical time series analysis and builds a bridge between the theories of dynamical systems, signal processing and graph theory.\\

\noindent In recent years, this mapping has been used to provide a topological characterization of different routes to low dimensional chaos \cite{jns, quasi, pre2013}, or different types of stochastic and chaotic dynamics \cite{nonlinearity}. From an applied angle, it is being widely used to extract in a simple and computationally efficient way informative features for the description and classification of empirical time series appearing in several areas of physics including optics \cite{physics3}, fluid dynamics \cite{fluiddyn0, fluiddyn1, fluiddyn2}, geophysics \cite{physics2} or astrophysics \cite{suyal, Zou}, and extend beyond physics in areas such as physiology \cite{physio1, meditation_VG}, neuroscience \cite{neuro} or finance \cite{ryan1} to cite only a few examples. Whenever each element in a given classification task is naturally encoded as an ordered sequence,
one can map such sequence into a visibility graph and subsequently extract a certain set of topological properties of these graphs as the feature vector with which to train classifiers in supervised learning tasks.\\

\noindent Here we propose to extend this methodology from time series $\{{\bf x}(t)\}_{t=1}^N$ to  scalar fields $h(x,y):\mathbb{R}^d\to \mathbb{R}$. This extension, which has only been scarcely explored \cite{rowcolumn} is conceptually closer to the original context of visibility graphs \cite{original} and enables the possibility of constructing the visibility graphs of images, landscapes, and general large-scale spatially-extended surfaces. In what follows we will introduce the concept along with a few definitions and properties. In section III we provide analytical results on some topological properties of these graphs
associated to some types of real-valued matrices which can be understood as the high and low disorder limits of real-valued scalar fields. In particular, we find a closed expression for the degree distribution of these graphs associated to uncorrelated random fields of generic dimension, extending the result known for one-dimensional time series. As this result holds independently of the field's marginal distribution, we show that this result directly yields a statistical randomness test, applicable in arbitrary dimensions. In section IV we showcase its usefulness by discriminating two-dimensional white noise from two-dimensional lattice of diffusively coupled chaotic maps (a system that generated high dimensional spatio-temporal chaos). In section V we discuss the range of potential applications of this combinatorial framework and we further briefly illustrate its usefulness for characterizing the process of oncogenesis.

\section{Definitions and basic properties}
\noindent {\bf Definition }(VG) {\it  Let ${\cal S}=\{x_1,\dots,x_N\}$ be a ordered sequence of $N$ real-valued, scalar datapoints. A Visibility Graph (VG) is an undirected graph of $n$ nodes, where each node $i\in [1,N]$ is labelled by the time order of its corresponding datum $x_i$. Hence $x_1$ is mapped into node $i=1$, $x_2$ into node $i=2$, and so on.
Then, two nodes $i$ and $j$ (assume $i<j$ without loss of generality) are connected by a link if and only if one can draw a straight line connecting $x_i$ and $x_j$ that does not intersect any intermediate datum $x_k, \ i<k<j$. Equivalently, $i$ and $j$ are connected if the following {\it convexity} criterion is fulfilled:}
$$x_k< x_i + \frac{k-i}{j-i}[x_j-x_i],\ \forall k: i<k<j$$
The same definition applies to a Horizontal Visibility Graph (HVG) but in this latter graph two nodes $i$, $j$ (assume $i<j$ without loss of generality) are connected by a link if and only if one can draw a {\it horizontal} line connecting $x_i$ and $x_j$ that does not intersect any intermediate datum $x_k, \ i<k<j$. Equivalently, $i$ and $j$ are connected if the following {\it ordering} criterion is fulfilled:\\
$$x_k<\inf(x_i,x_j),\ \forall k: i<k<j$$
From a combinatoric point of view, HVGs are outerplanar graphs with a Hamiltonian path \cite{severini}, i.e. noncrossing graphs as defined in algebraic combinatorics \cite{flajo}.
Note that the former definitions focus on {\it discrete} sequences, such that the index labeling is such that $i+1\equiv i+\Delta$, where $\Delta$ is the spacing between data. Interestingly, both VG and HVG are invariant under changes in $\Delta$. In particular, this enables to directly consider the {\it continuous} version of a discrete time series simply as the limit $\Delta \to 0$. This invariance property will allow treating continous scalar fields as the $\Delta \to 0$ limit of matrices as we will show later.\\  

\begin{figure}
\centering
\includegraphics[width=0.45\columnwidth]{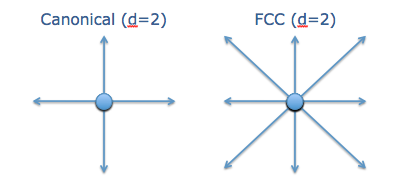}
\caption{Illustration of the two extension classes (canonical and FCC) for $d=2$.}
\label{possibilities}
\end{figure}

\noindent {\bf Extension classes. }One can now extend the definition of visibility to handle two-dimensional manifolds, by simply extending the visibility criteria along one-dimensional sections of the manifold. The question is, in how many different ways one can do that? As a matter of fact, there exist several possibilities, here we consider just a few of them. We firstly consider manifolds of dimension $d$ which have a natural Euclidean embedding and define two extension classes, labelled as {\it canonical} and {\it FCC} respectively. In the canonical extension class, the rule of thumb for extending the definition of a visibility graph to a manifold of dimension $d$ will be by applying the VG/HVG to $d$ orthogonal sections of the manifold (which define $n=2d$ directions). In other words, at each point of the manifold one constructs the VG/HVG in the direction of the (canonical) Cartesian axis. On the other hand, the FCC extension class allows an additional number of sections in the direction of the main diagonals. Accordingly, in this second class the number of directions is $n=2d+2^d$ directions (see figure \ref{possibilities} for an illustration in the case $d=2$). Finally, a third extension class (which in this work will only be studied for $d=2$ flat surfaces) is defined by taking $n$ directions in such a way that the set of $n$ vectors make an homogeneous angular partition of the plane with constant angle $2\pi/n$. This class is labelled as the order-$n$ class. Obviously, the order-$8$ and order-$4$ classes coincide, when $d=2$, with the FCC and canonical classes respectively. These special classes are indeed of special relevance as they are the most natural algorithmic implementation for image processing \cite{image_processing}. We are now ready to give a more formal definition of visibility graphs in these extension classes.\\

\noindent {\bf Definition} (IVG$_n$) {\it Let ${\cal I}$ be a $N\times N$ matrix, where ${\cal I}_{ij}\in \mathbb{R}$ and $N>0$. For an arbitrary entry $ij$, make an angular partition of the plane into $n$ directions, such that direction labelled as $p$ makes an angle with the row axis of $2\pi(p-1)/n$. The {\it Image Visibility Graph of order n} $\text{IVG}_n$ is a graph with $N^2$ nodes, where each node is labelled by a duple $ij$ in association with the indices of the entry ${\cal I}_{ij}$, such that two nodes $ij$ and $i'j'$ are linked if}
\begin{enumerate}
\item {\it $i'j'$ belongs to one of the $n$ angular partition lines, and}
\item {\it ${\cal I}_{ij}$ and ${\cal I}_{i'j'}$ are linked in the VG defined over the ordered sequence which includes $ij$ and $i'j'$.}
\end{enumerate}
The {\it Image Horizontal Visibility Graph} ($\text{IHVG}_n$) follows equivalently if in the second condition we make use of HVG instead of VG. Note that in the preceding definition, ${\cal I}$ can be understood as a two-dimensional square lattice, which is naturally embedded in $\mathbb{R}^2$ if we associate a certain lattice length $\Delta_p>0$ to the separation between any two neighbors in each direction $p$. In the limit $N\to \infty, \Delta_p\to 0$ this matrix ${\cal I}$ converges in some mathematically well-defined sense to a continuous scalar field $h(x,y):\mathbb{R}^2\to R$. Accordingly, the continuous version of these graphs can be obtained for $n\to \infty$, and in that case I(H)VG$_\infty$ would be an infinite graph. In this work we keep $n$ finite and from now on only consider finite discretizations of scalar fields, however the infinite case is certainly of theoretical interest and is left for future investigations.\\

\noindent For a given dimension $d$, one can define in a similar fashion the Visibility Graphs in the canonical extension class labelled IVG$^c$(d) by modifying condition (1): $i'j'$ belongs to one of the $d$ Cartesian axis which span $\mathbb{R}^d$ and have origin in $ij$. Analogously, the Visibility Graphs in the FCC extension class IVG$^{FCC}$(d) are obtained by modifying again condition (1) appropriately to allow visibility in the main diagonals. Finally, again the Horizontal version follows equivalently if in the second condition we make use of HVG instead of VG.\\

\noindent A trivial but important remark is that $\forall {\cal I}$, I(H)VG$_4$=I(H)VG$^c$(2) and I(H)VG$_8$=I(H)VG$^{FCC}$(2). Note also that the special class IVG$^c$(2) has been explored recently under the name row-column visibility graph \cite{rowcolumn}.\\
Once any of these graphs has been extracted from a given matrix ${\cal I}$, one can further compute standard topological properties on this graph using classical measures from Graph Theory \cite{GT} or recent metrics defined in Network Science \cite{NS}, which in turn might be used to provide a topological characterization of ${\cal I}$. For instance, the degree $k$ of a node is the number of links of that node. This allows to construct the degree matrix ${\bf K} \in {\mathbb{N}^{N\times N}}$, where ${\bf K}_{ij}$ is the degree of node labelled with the pair $i,j$. The degree distribution $P(k)$ determines the probability of finding a node of degree $k$ and can be straightforwardly computed from the degree matrix. In this work for concreteness we will only consider these metrics, however we should emphasize here that a large toolbox of measures could be used for feature extraction in context-dependent applications. Here we are motivated to use these very simple metrics as it has recently been proved that, in the one-dimensional case, the set of degrees is on bijection with the adjacency matrix and hence is indeed an optimal feature \cite{Luque_Theorem}.\\

\noindent In what follows we depict some exact results on the topology of these graphs associated to simple types of matrices which can be understood as the high order and high disorder limits of real images. From now on we only consider the Horizontal version of the visibility criteria, and we assume $N\to \infty$ to avoid border effects.

\section{Some exact results}
\noindent {\bf Periodicity: monochromatic images and chess. }
We start by considering trivial configurations at the end of total order. For monochromatic images where ${\cal I}_{ij}=c$, the IHVG$_n$ is such that ${\bf K}_{ij}=n$ and thus $P(n)=1$ and $P(k\neq n)=0$. Then we can consider chess. This is a periodic lattice, where in each row the same periodic sequence is represented (black,white,black,\dots)$\equiv (1,-1,1,-1,\dots)$, except for a one-step translation in even rows. Accordingly, neglecting boundary conditions ${\cal I}_{ij}= 1$ if $i \cdot j$ is odd and $-1$ otherwise. For IHVG$_4$ we find ${\bf K}_{ij}=8$ if $i \cdot j$ is odd and $4$ otherwise. For IHVG$_8$ we find ${\bf K}_{ij}=12$ if $i \cdot j$ is odd and $8$ otherwise. From this latter matrix the degree distribution is simply $P(k)=1/2$ for $k=8,12$ and zero otherwise. For other types of periodic structures it is easy to see that the degree matrix will inherit such periodicity and thus the degree distribution will only be composed by a finite number $q$ of non-null probabilities, where $q$ in turn is typically bounded by a function that depends on the period of the periodic structure.\\ 

\noindent {\bf Uncorrelated random fields. }
We then consider a limit configuration at the end of total disorder: a two-dimensional uncorrelated random field, i.e. white noise. Then, the following theorem holds for the degree distribution of IHVG$_n$:\\

\noindent{{\bf Theorem. } {\it Consider an $N \times N$ matrix with entries ${\cal I}_{ij}=\xi$, where $\xi$ is a random variable sampled from a distribution $f(x)$ with continuous real support $x\in(a,b)$.  Then, for $n>0$ and in the limit $N\to \infty$ the degree distribution of the associated $\text{IHVG}$$_n$ converges to}
\begin{equation}
P(k)=     \begin{cases}
      \big(\frac{1}{n+1}\big)\big(\frac{n}{n+1}\big)^{k-n}, & \text{if}\ k\geq n \\
      0, & \text{otherwise}
    \end{cases}
    \label{theorem}
\end{equation}
\noindent A few comments are in order before presenting a proof is presented. First, note that this equation reduces, for $n=2$ ($d=1$), to the well-known result for time series of i.i.d. variables $P(k)=(1/3)(2/3)^{k-2}$ \cite{PRE}. Second, in the specific class $n=8$ (equivalent to the FCC class in $d=2$, being this the selected version for image processing \cite{image_processing}), eq.\ref{theorem} yields
\begin{equation}
P(k)=     \begin{cases}
      \big(\frac{1}{9}\big)\big(\frac{8}{9}\big)^{k-8}, & \text{if}\ k\geq 8 \\
      0, & \text{otherwise}
    \end{cases}
\label{n8}
\end{equation}
\noindent Third, note that in the limit of large $n$ we would have a continuous visibility scanning. The extension for any generic $n$ can also be directly interpreted as a generalization to higher dimensional (discrete) scalar fields, so it is easy to show that eq.\ref{theorem} also applies to the degree distribution of (i) the canonical extension for dimension $d=n/2$ (i.e. only even values of $n$ are allowed in this case), and (ii) the FCC extension for dimension $d$, where $n=2d+2^d$ (i.e. for $n=8,14,24,42,\dots$). We are now ready to provide the proof of the theorem.\\

\noindent {\bf Proof. } The proof essentially makes use of the diagrammatic formalism introduced in \cite{PRE, nonlinearity} where, in the case of time series, the probability of each degree was expanded in a series of terms, each term associated to a different diagram and contributing with different amplitude.\\
Let us start by considering the concrete case $n=8$ (which describes the case implemented in our algorithm for image filtering) and we will generalize for all $n$ thereafter. Using the jargon developed in \cite{PRE, nonlinearity}, a node chosen at random which has horizontal visibility of $k$ others can be modeled as a {\it seed} (contributing with probability $\mathfrak{S}$) which has visibility of $k-8$ {\it inner} nodes (contributing with $\mathfrak{I}$) distributed along the $n=8$ directions (such that direction $i$ contributes with $k_i$ inner nodes),
and whose visibility is finally bounded by 8 {\it bounding} nodes (contributing with probability $\mathfrak{B}$). 
The probability of this event can thus be formally expressed as
\begin{equation}
P(k)=\sum_{\{k_1,k_2\dots k_8\}}{\mathfrak S}{\mathfrak B}^8\prod_{i=1}^8{\mathfrak I}_{k_i},
\label{inicial}
\end{equation}
where the sum enumerates all admissible combinations of $\{(k_1,k_2,\dots,k_8)\}$ such that $\sum_{i=1}^8=k-8$ (by construction, every node always has visibility of its boundary, here formed by $n=8$ nodes). It is easy to see that a possible enumeration is 
$$\  k_i=0,1,\dots,k-8-\sum_{m=1}^{i-1}k_m \ \text{for} \ i=1,2,\dots,7;\  k_8=k-8-\sum_{i=1}^7 k_i.$$
Making use of the cumulative distribution  $F(x)=\int_a^x f(x')dx'$ (with $F(a)=0, \ F(b)=1$) and following \cite{PRE, nonlinearity}, geometrically it is easy to see that
$${\mathfrak S}=\int_a^b f(x_0)dx_0; \ 
{\mathfrak B}=\int_{x_0}^b f(x)dx=1-F(x_0); \ 
$$
To describe the probability of finding $p$ inner nodes $\mathfrak{I}_p$, by construction we shall take into account that an arbitrary number $r$ (from zero to an infinite amount) of {\it hidden} data can lie in between every pair of aligned inner nodes. Such arbitrary number of hidden data should contribute with the following amplitude
$$\sum_{r=0}^\infty \prod_{j=1}^r \int_a^x f(n_j)dn_j=\frac{1}{1-F(x)},$$
where we have used the properties of the cumulative distribution to find the last identity. Accordingly, the concatenation of $p$ inner data which might have an arbitrary number of interspersed hidden data can be expressed as

\begin{equation}
\mathfrak{I}_p=\int_a^{x_0}
\frac{f(x_1)dx_1}{1-F(x_1)}\prod_{j=1}^{n-1}\int_{x_j}^{x_0}\frac{f(x_{j+1})dx_{j+1}}{1-F(x_{j+1})}.\label{recurrencia}
\end{equation}
This latter calculation is easy but quite tedious. One proceeds
to integrate equation \ref{recurrencia} step by step and a recurrence quickly becomes evident.
One can easily prove by induction that
$${\mathfrak I}_p=\frac{(-1)^p}{p!}[\ln(1-F(x_0))]^p.$$
We are thus ready to tackle eq. \ref{inicial}.
Taking advantage of the closure $\sum_{i=1}^8 K_i=k-8$, we first have
$$\prod_{i=1}^8{\mathfrak I}_{k_i}=\frac{(-1)^{k-8}[\ln(1-F(x_0))]^{k-8}}{\prod_{i=1}^8(k_i)!},$$
so after some reordering, 
$$P(k)=\sum_{\{k_1,k_2\dots k_8\}}\frac{(-1)^{k-8}}{\prod_{i=1}^8(k_i)!}\int_a^b f(x_0)(1-F(x_0))^8[\ln(1-F(x_0))]^{k-8}dx_0.$$
Now, in this latter equation the integral is easy to compute:
$$\int_a^b f(x_0)(1-F(x_0))^8[\ln(1-F(x_0))]^{k-8}dx_0=(-1)^{k-8}(k-8)!\bigg(\frac{1}{9}\bigg)^{k-7}$$
\noindent Consider finally the term 
\begin{eqnarray}
\sum_{\{k_1,k_2\dots k_8\}}\frac{(k-8)!}{\prod_{i=1}^8(k_i)!}=(k-8)!\sum_{k_1=0}^{k-8}\sum_{k_2=0}^{k-8-k_1}\dots\sum_{k_7=0}^{k-8-\sum_{j=1}^7 k_j}\frac{1}{k_1!}\frac{1}{k_2!}\dots\frac{1}{k_7!}\frac{1}{(k-8-\sum_{j=1}^7 k_j)!}=8^{k-8}
\end{eqnarray}
where the last indentity was found by iteratively applying the binomial theorem
$\sum_{k=0}^{a} {a \choose k} r^{k}=(1+r)^{a}.$ 
Altogether, we can write down explicitly for $n=8$
$$P(k)=\bigg(\frac{1}{9}\bigg)\bigg(\frac{8}{9}\bigg)^{k-8}$$
for $k\geq 8$ and zero otherwise. This result is independent of $f(x)$ as expected since HVG is an order statistic \cite{nonstationary}, and coincides with eq. \ref{theorem} for $n=8$ (i.e. eq. \ref{n8}).\\ 

\noindent We are now ready to generalize the whole derivation. For a generic $n$, trivially
$$P(k)=\sum_{\{k_1,k_2\dots k_n\}}\frac{(-1)^{k-n}}{\prod_{i=1}^n(k_i)!}\int_a^b f(x_0)(1-F(x_0))^n[\ln(1-F(x_0))]^{k-n}dx_0.$$
with
$$\int_a^b f(x_0)(1-F(x_0))^n[\ln(1-F(x_0))]^{k-n}dx_0=\bigg(\frac{1}{n+1}\bigg)^{k-n+1}(-1)^{k-n}(k-n)!$$
such that
$$P(k)=\bigg(\frac{1}{n+1}\bigg)^{k-n+1}\sum_{\{k_1,k_2\dots k_n\}}\frac{(k-n)!}{\prod_{i=1}^n(k_i)!}.$$
Finally since
$$\sum_{\{k_1,k_2\dots k_n\}}\frac{(k-n)!}{\prod_{i=1}^n(k_i)!}=n^{k-n},$$
we find 
$$P(k)=\bigg(\frac{1}{n+1}\bigg)^{k-n+1}n^{k-n}=\bigg(\frac{1}{n+1}\bigg)\bigg(\frac{n}{n+1}\bigg)^{k-n},$$
what concludes the proof. $\blacksquare$ \\

\noindent Note that a similar result can be found much easily at the expense of using a non-rigorous heuristic argument. In the case $n=8$, the probability that the seed node has visibility of exactly $k$ nodes can be expressed as the probability that there are $k-8$ nodes that are not bounding times the probability that after these, the boundary prevents larger visibility. Accordingly, we shall write
$$P(k)=(1-P(8))^{k-8}P(8)$$
For $k=8$, $k_i$ only take the value $k_i=0 \ \forall i=1\dots 8$, hence this term is straightforward to compute
$$P(8)={\mathfrak S}{\mathfrak B}^8=\int_a^b f(x_0)\bigg[\int_{x_0}^b f(x)dx\bigg]^8 dx_0=\frac{1}{9}, \ \forall f$$
which then yields the correct shape for $P(k)$:
$$P(k)=(1-P(8))^{k-8}P(8)=\bigg(\frac{1}{9}\bigg)\bigg(\frac{8}{9}\bigg)^{k-8}$$
A similar argument can be used for a generic $n$, yielding
$$P(k)=(1-P(n))^{k-n}P(n)=\bigg(\frac{1}{n+1}\bigg)\bigg(\frac{n}{n+1}\bigg)^{k-n}$$
for $k\geq n$ and zero otherwise, in good agreement with eq. \ref{theorem}.\\

\begin{figure}
\centering
\includegraphics[width=0.9\columnwidth]{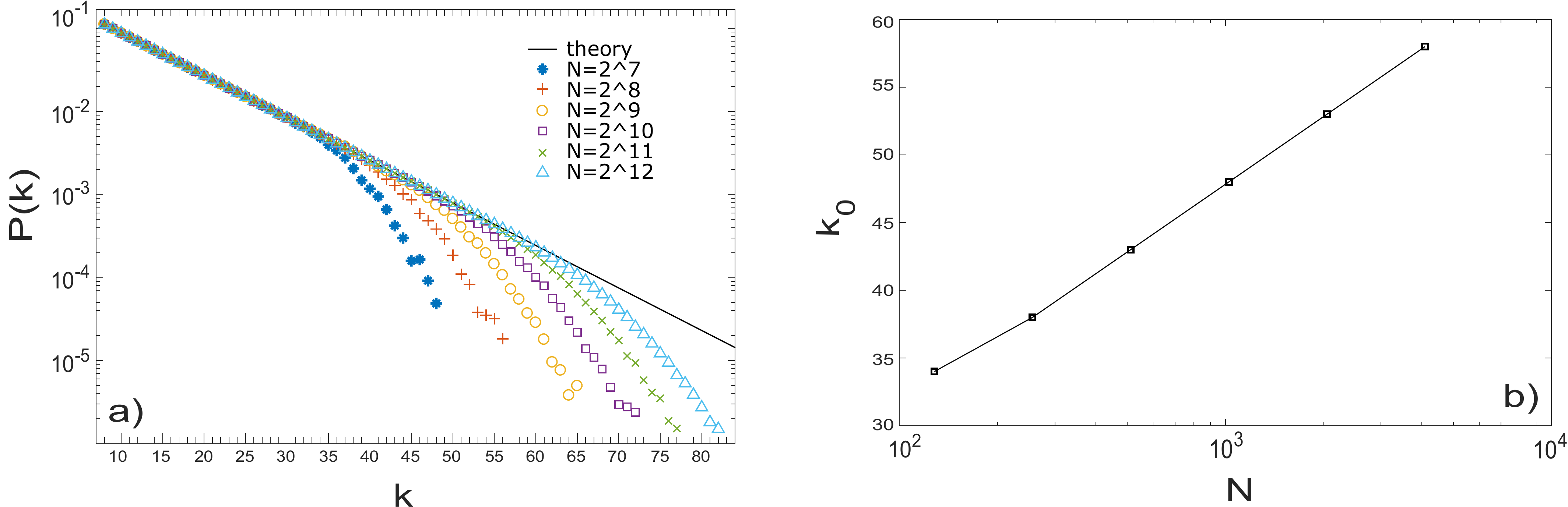}
\caption{{\it (Left, Panel a)} Semi-log plot of the degree distribution (ensemble averaged over 10 realizations) of IHVG$_8$ associated to $N\times N$ matrices with i.i.d. uniform U[0,1]random entries, for $N=2^7,2^8,\dots,2^{12}$. The solid line is the theoretical value of $P(k)$ given by eq.\ref{theorem} for $n=8$. In every case we find excellent agreement for $k<k_0$, where $k_0$ is a cut-off value that denotes the onset of finite size effects.
{\it (Right Panel b)} Linear-log plot of the cut-off $k_0$ as a function of the system's size $N$ for the same data of Panel a, suggesting a logarithmic scaling $k_0\sim c\log N$.}
\label{Pk_N}
\end{figure}

\noindent {\bf Finite size effects. }To assess the convergence speed to eq. \ref{theorem} for finite $N$, we have estimated the degree distribution of IHVG$_8$ associated to $N\times N$ random matrices whose entries are i.i.d. uniform random variables U[0,1]. In figure \ref{Pk_N} we plot, in semi-log scales, the resulting (finite size) degree distributions, for different $N=2^7,2^8,\dots,2^{12}$. As we can see, the distributions are on excellent agreement with eq. \ref{theorem} for $k\leq k_0$, where the location of the cut-off value $k_0$ scales logarithmically with the system's size $N$ as shown in the bottom of the figure. In other words, finite size effects only affects the tail of the distribution, which converges logarithmically fast with $N$.

\section{A simple application}
The results for uncorrelated random fields found in the previous section are indeed of practical interest because eq.\ref{theorem} holds independently of the noise marginal distribution $f$. Resorting to the contrapositive, if the degree distribution of IHVG$_n$ deviates from eq.\ref{theorem} for some empirical field $\cal I$, one can conclude that the field is not uncorrelated noise. This theorem thereby allows for the straightforward design of a randomness statistical test which would be applicable to data structure of arbitrary dimension $d$, where $n(d)=2d$ if one uses the canonical extension class, or $n(d)=2d+2^d$ in the case of FCC.\\

\begin{figure}
\centering
\includegraphics[width=0.7\columnwidth]{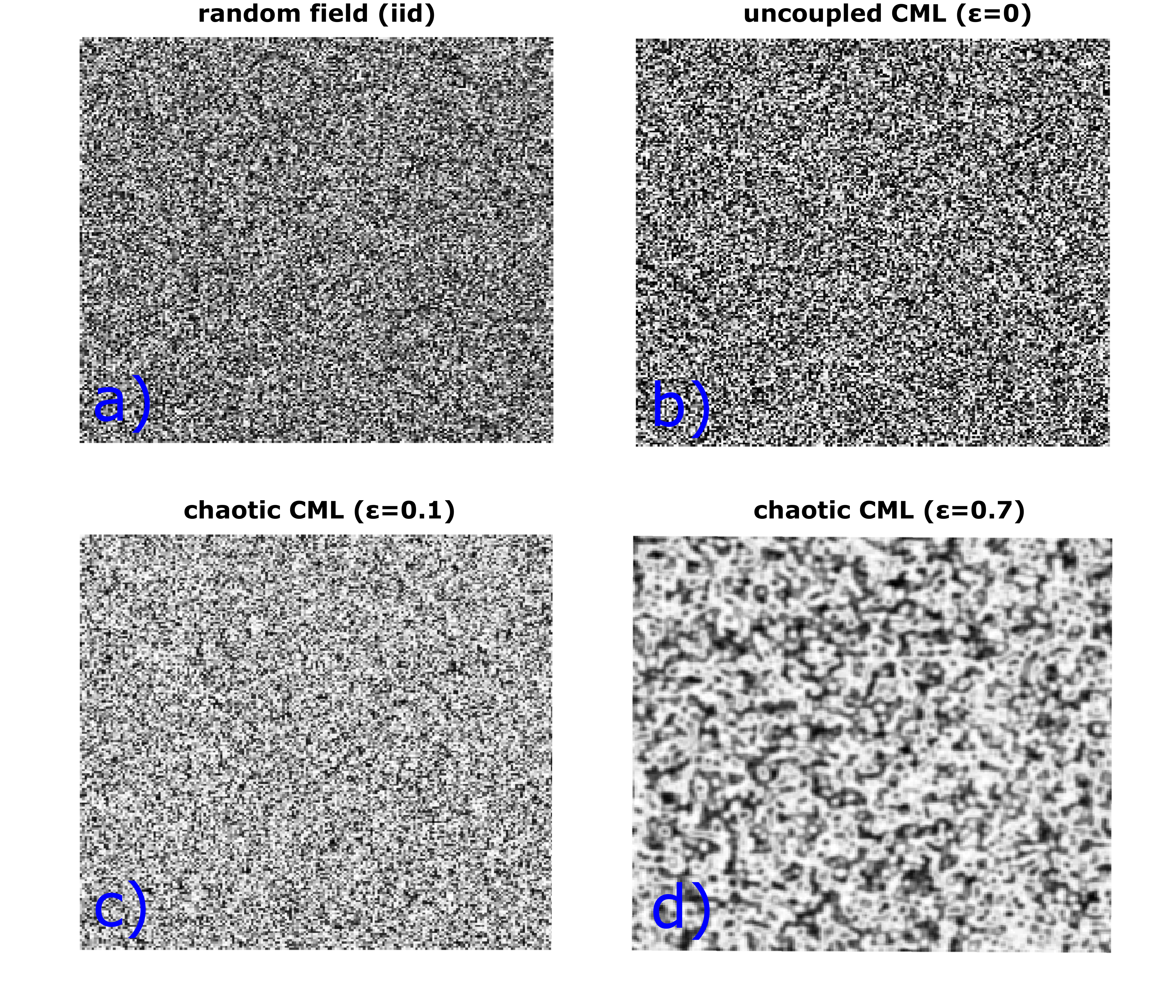}
\caption{Grayscale plots of 200x200 matrices describing: (top, left panel) i.i.d. uniform U[0,1] random variables (uniform white noise); (top, right panel) a snapshot of a two-dimensional lattice of diffusively coupled chaotic logistic maps with coupling strength $\epsilon=0$ (effectivey being uncoupled and therefore a snapshot of uncorrelated Beta distributed white noise); (bottom, left panel) the same coupled map lattice for weak coupling $\epsilon=0.1$ for which the system displays fully-developed turbulence (a state of spatio-temporal chaos with a high dimensional attractor). 
(bottom, right panel) the same coupled map lattice for strong coupling $\epsilon=0.7$. In this latter case the system shows strong spatial correlations and is easily distinguishable from the rest.}
\label{iid_vs_CML}
\end{figure}
\noindent {\bf Coupled Map Lattices. }To illustrate this we consider a simple application of discriminating noise from high-dimensional chaos. 
Chaotic processes display irregular and unpredictable behavior which is often confounded with randomness, however chaos is a deterministic process which indeed hides in some cases some patterns that can be extracted by appropriate techniques. The endeavor of distinguishing noise from chaos has been an area of intense research activity in the last decades \cite{libro_chaos} and applications have pervaded nearly every scientific discipline where complex, irregular empirical signals emerge. Here we consider spatially extended structures and thus we will be dealing with spatio-temporal chaos, i.e. chaotic behavior in space and in time, and we will explore whether if visibility graphs are able to distinguish such dynamics from simple randomness.
Let us define ${\cal I}(t)$ as a two dimensional square lattice of $N^2$ diffusively coupled chaotic maps which evolve in time \cite{CML}. In each vertex of this coupled map lattice (CML) we allocate a fully chaotic logistic map $x_{t+1}=Q(x_t),\ Q(x)=4x(1-x)$, and the system is then spatially coupled as it follows:
\begin{equation}
{\cal I}_{ij}(t+1)=(1-\epsilon)Q[{\cal I}_{ij}(t)] + \frac{\epsilon}{4}\sum_{i',j'}Q[{\cal I}_{i'j'}(t)],
\label{CMLeq}
\end{equation}
where the sum extends to the Von Neumann neighborhood of $ij$ (four adjacent neighbors)
The update is parallel and we use periodic boundary conditions. The coupling strength $\epsilon \in[0,1]$. For $\epsilon=0$ the system is uncoupled and the $N^2$ logistic maps evolve independently. For positive $\epsilon>0$ there is a balance between the internal (chaotic) dynamics which drives a local tendency towards inhomogeneity and the diffusion term (in the right
hand of the equation one can easily recognize the discrete version of the Laplacian) which induces a global tendency
towards homogeneity in space. This balance is tuned by $\epsilon$, acting as an effective viscosity constant, and the system evolves into different spatio-temporal dynamics as $\epsilon$ varies. For a small yet positive value of the coupling the system displays so-called Fully Developed Turbulence, a phase with incoherent spatiotemporal
chaos and high dimensional attractor \cite{CML}. In other words, the system evolves both temporally and spatially in a very irregular way, yet it is not totally uncorrelated. For illustration, in figure \ref{iid_vs_CML} we plot, for $N=200$, grayscale {\it snapshots} of this system for $\epsilon=0$ (uncoupled), $\epsilon=0.1$ (weak coupling) and $\epsilon=0.7$ (strong coupling) along with a $200\times 200$ matrix of $U[0,1]$ i.i.d. random variables (white noise). Note that the snapshot of the uncoupled case reduces to a collection of independent and identically distributed chaotic variables with a marginal distribution that coincides with the invariant measure of the fully chaotic logistic map: the Beta distribution $\mathbb{B}(1/2,1/2)=\pi^{-1}x^{-1/2}(1-x)^{-1/2}$. In other words, such a snapshot is indistinguishable from white, Beta-distributed noise, which should be then equivalent under the IHVG mapping to any type of white noise and should therefore fulfill our theorem. When $\epsilon>0$ spatial correlations settle in and the snapshots are in theory statistically different, however this difference is only evident for large coupling.\\

\begin{figure}
\centering
\includegraphics[width=0.48\columnwidth]{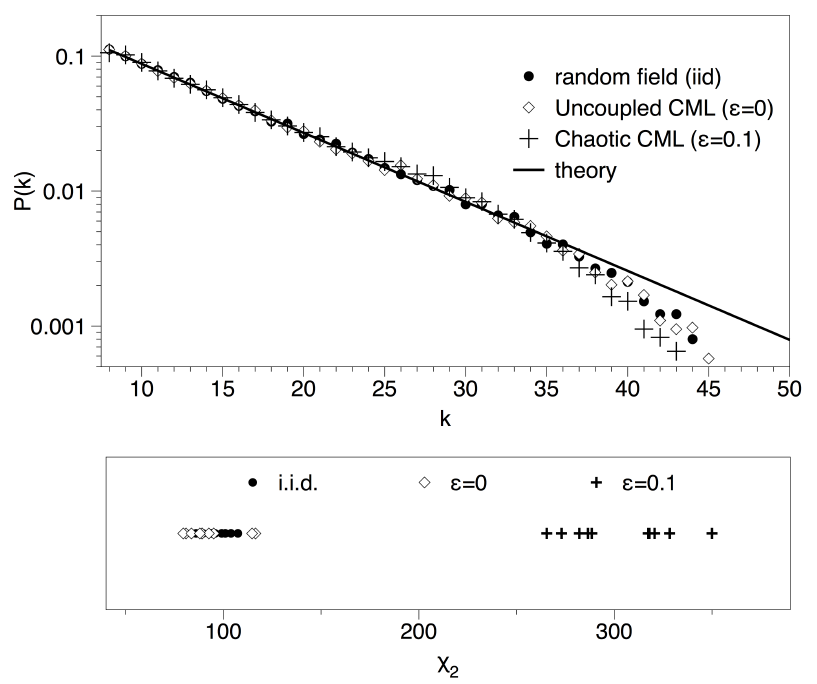}
\includegraphics[width=0.5\columnwidth]{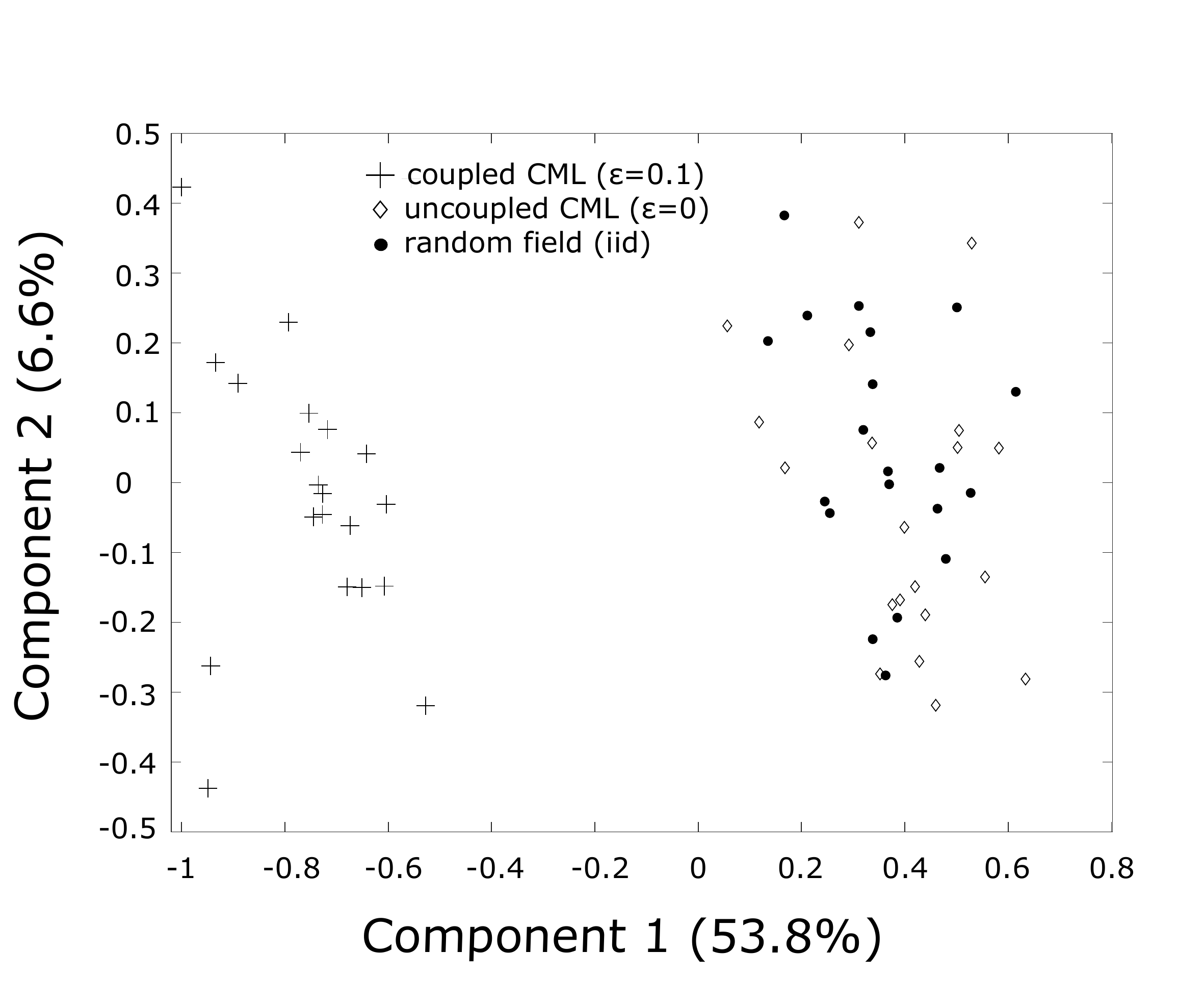}
\caption{({\it Left Panel}) Semi-log plot of the degree distribution of IHVG$_8$ associated to a two-dimensional uncorrelated random field of uniform random variables (black dots), and two-dimensional coupled map lattices of diffusively coupled fully chaotic logistic maps, for coupling constant $\epsilon=0$ (diamonds) and $\epsilon=0.1$ (crosses). The solid line is eq.\ref{theorem} for $n=8$. Deviations from the exponential law in the tail are due to finite size effects (in every case matrices are $200\times200$). Note that the $\epsilon=0$ case is effectively a spatially uncorrelated field with i.i.d. entries (with marginal distribution equivalent to the invariant measure of an isolated logistic map, i.e. the beta distribution). For $\epsilon=0.1$ the system is weakly coupled and displays fully-developed turbulence (spatio-temporal chaos with high-dimensional attractor, i.e. the snapshot is weakly correlated). All three snapshots (a,b,c in figure \ref{iid_vs_CML}) look very similar and, expectedly, they all display {\it apparently similar} degree distributions. 
({\it Bottom, left panel}) Here we consider 20 realizations of each of the three systems, and in each case compute the $\chi_2$ statistic (see the text) measuring the deviation of the empirical degree distribution ($k<44$) from the theory for random fields. As expected, the i.i.d. cases (random field and snapshot of the uncoupled logistic maps) are indistinguishable, but the weakly coupled system is clearly distinguished, finding stronger deviations from eq. \ref{theorem} than those found due to finite size effects. ({\it Right panel}) Principal component analysis (PCA) of the set of degree distributions for the 60 realizations explored in the left (bottom) panel. Each degree distribution $P(k)$ has been projected in a two-dimensional space spanned by the first two principal components (this subspace accounts for 60$\%$ of the variability). One does not need to apply any clustering algorithm as the non-random matrices are very clearly clustered together and apart from the i.i.d. cases.}
\label{Pkchi2}
\end{figure}

\noindent {\bf Distinguishing noise from chaos. }To explore such differences we can exploit our theorem as it follows: first, we estimate the degree distribution of the IHVG$_8$ of each snapshot, and compare against the theoretical equation for white noise. To account for finite size effects, it is necessary to compare the estimation of the chaotic case not just with eq.\ref{theorem} but also with a finite i.i.d. sample. 
We have generated 20 realizations of each process (random uniform noise, $\epsilon=0$ and $0.1$) and have extracted the degree distribution of IHVG$_8$ for each case. Sample results of these distributions can be shown in the left panel of figure \ref{Pkchi2} along with the theoretical prediction for i.i.d (eq. \ref{n8}). As expected, the distributions are {\it apparently} very well approximated by eq. \ref{n8} in every case (there are strong deviations for $k>35$ but this is due to finite size effects as similar deviations take place for the i.i.d. white uniform noise case). To quantify potential deviations from the theory (which according to the theorem would imply non-randomness), for each case we have computed the $\chi^2$ statistic $$\chi^2=N\sum_k \frac{[P_{\text{th}}(k)-P_{\text{exp}}(k)]^2}{P_{\text{th}}(k)},$$
where we have taken $k=8,9,\dots,44$. Results are shown in the bottom panel of figure \ref{Pkchi2}, showing now a clear separation between the uncorrelated cases (uncoupled chaotic maps and uniform white noise) and the weakly coupled system. This clear distinction is further confirmed in a principal component analysis (PCA) depicted in the right panel of the same figure, where each degree distribution $P(k)$ has been projected in a two-dimensional space spanned by the first two principal components (this subspace accounts for 60$\%$ of the variability). One does not need to apply any clustering algorithm as the non-random matrices are very clearly clustered together and apart from the i.i.d. cases.\\

\begin{figure}
\centering
\includegraphics[width=0.53\columnwidth]{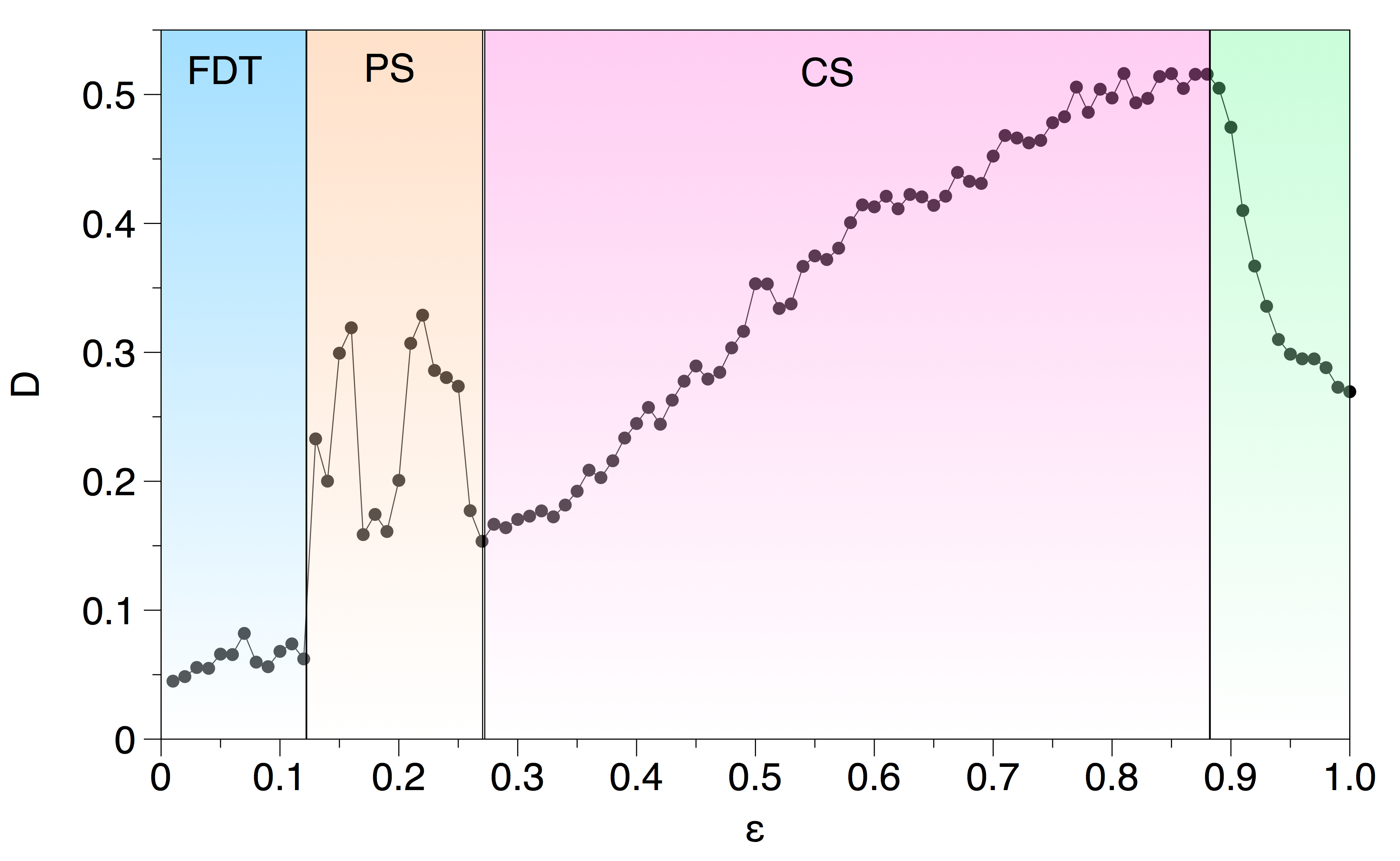}
\includegraphics[width=0.45\columnwidth]{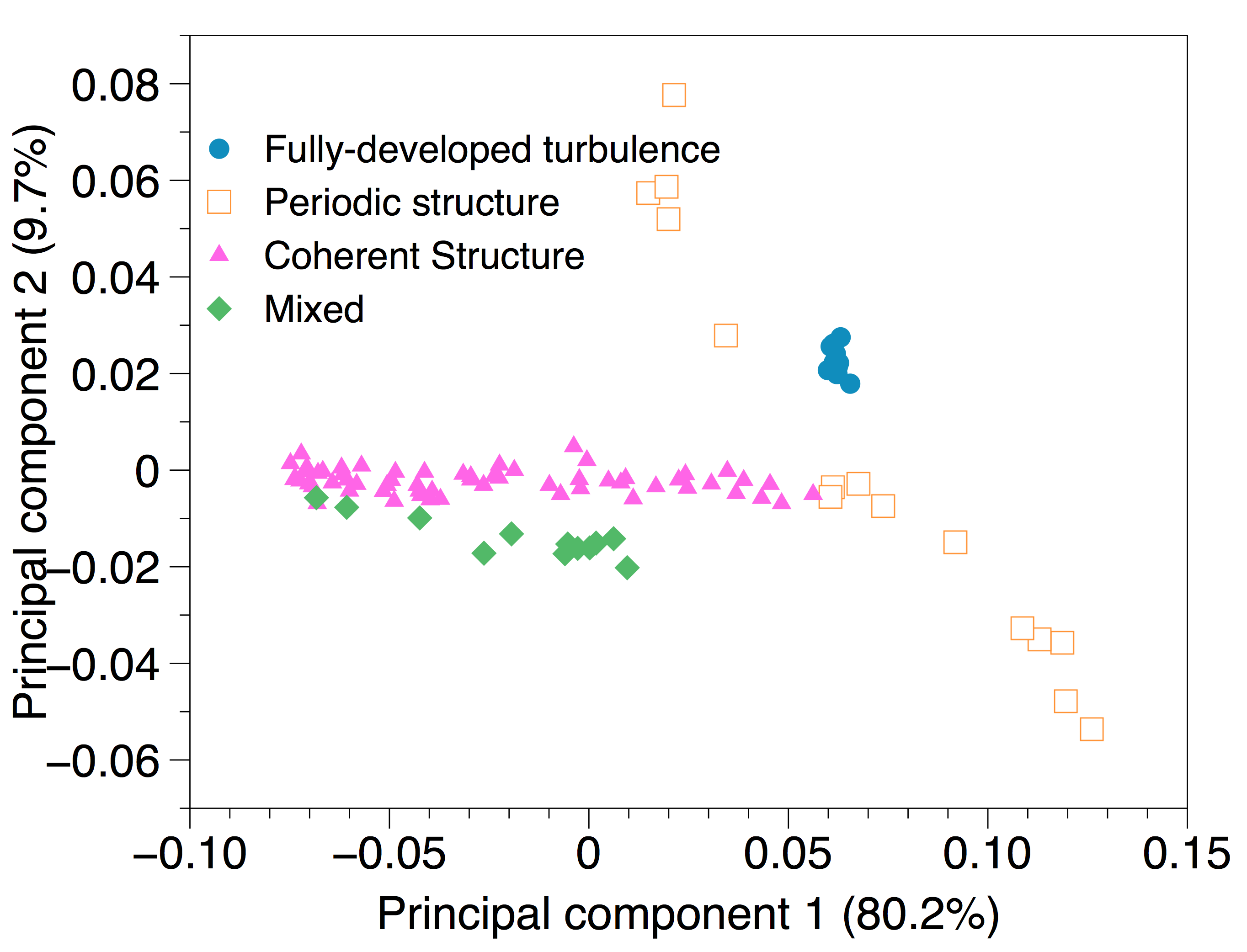}
\caption{{\it (Left panel)} Scalar parameter $D$ (see the text) as a function of the coupling constant $\epsilon$, compute from the degree distribution of IHVG$_8$ associated to $100 \times 100$ CMLs of fully chaotic logistic maps. $D$ captures the spatio-temporal phases: Fully-Developed Turbulence (FDT), Periodic Structure (PS), Coherent Structure (CS) and a mixed phase. Snapshots characteristic of these phases are depicted in figure \ref{CML_epsilon} in an appendix. {\it (Right panel)} Principal Component Analysis of the degree distributions of IHVG$_8$ associated to the same data of the left panel. The plot is a projection into the first two principal components (accumulating over 90$\%$ of the data variability). The different heuristic phases are highlighted.}
\label{phase_diagram}
\end{figure}

\noindent {\bf Phase diagram. } As mentioned previously, the spatio-temporal dynamics of the coupled map lattice show a rich phase diagram as we increase the coupling constant $\epsilon$. An easy way of encapsulating and visualizing such richness in a single diagram is presented in the left panel of figure \ref{phase_diagram}. For each $\epsilon$, we compute the degree distribution of the associated IHVG$_8$. Then we compute the distance $D$ between the degree distribution at $\epsilon$ and the corresponding result for $\epsilon=0$ (eq. \ref{n8}) $D=\sum_k |P(k)-(1/9)(8/9)^{k-8}|.$
$D$ acts as a scalar order parameter describing the spatial configuration of the CML, and interestingly, evidences sharp changes for the different phases, such as: for $\epsilon<0.12$, the system develops Fully-Developed Turbulence (FDT) with weak spatial correlations. This regime shifts to a Periodic Structure (PS) for $0.12<\epsilon<0.27$. This regime then parsimoniously shifts into a phase with spatially Coherent Structures (CS), which ultimately break down for $\epsilon>0.88$ in favor of periodic patterns. For $0.88<\epsilon<1$ the spatial structure shows a mix between CS and PS. We conclude that the degree distribution of the IHVG$_8$ captures this rich spatial structure, something confirmed via principal component analysis in the right panel of figure \ref{phase_diagram}.

\section{Discussion}
This framework allows the possibility of describing discretized scalar fields of arbitrary origin in a combinatorially compact fashion, and enables using the tools of graph theory and network science for the practical description and classification of spatially-extended data structures. For the sake of exposition and concreteness, in this work we have only used a couple of graph measures (degree matrix and degree distribution) which can be argued that were optimal in the one-dimensional case \cite{Luque_Theorem}, but it should be highlighted that this method is much more general and allows to extract from these graphs any desired property.\\
\noindent For $d=1$ the method was naturally designed for the task of time series analysis, and has been exploited accordingly and extensively in the last years -both from a theoretical point of view and for applications- as was acknowledged in the introduction section. Here we have presented a natural extension of these algorithms to deal with (discretized) scalar fields of arbitrary dimension, along with a few exact results on simple -yet relevant- cases. From a mathematical point of view, the task of characterizing the graphs in these extension classes provide a wide range of challenging open questions, which could parallel recent advancements in the one-dimensional case \cite{nonlinearity}. Now, what are the potential {\it applications} of this framework?\\

\begin{figure}[h]
\centering
\includegraphics[width=0.3\columnwidth]{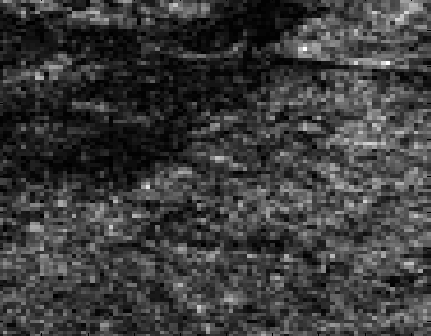}
\includegraphics[width=0.3\columnwidth]{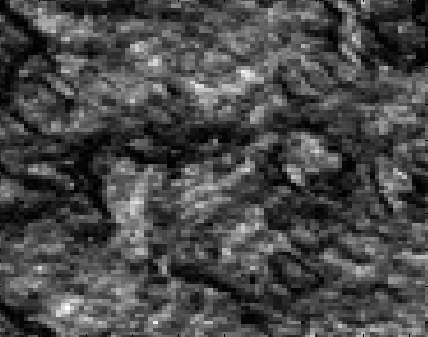}
\includegraphics[width=0.3\columnwidth]{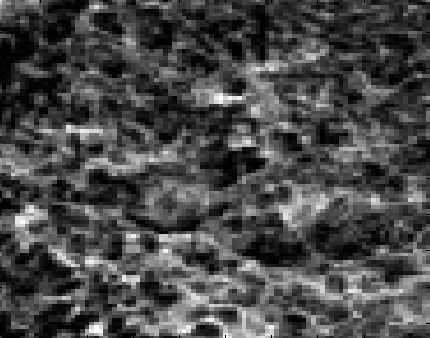}
\includegraphics[width=0.45\columnwidth]{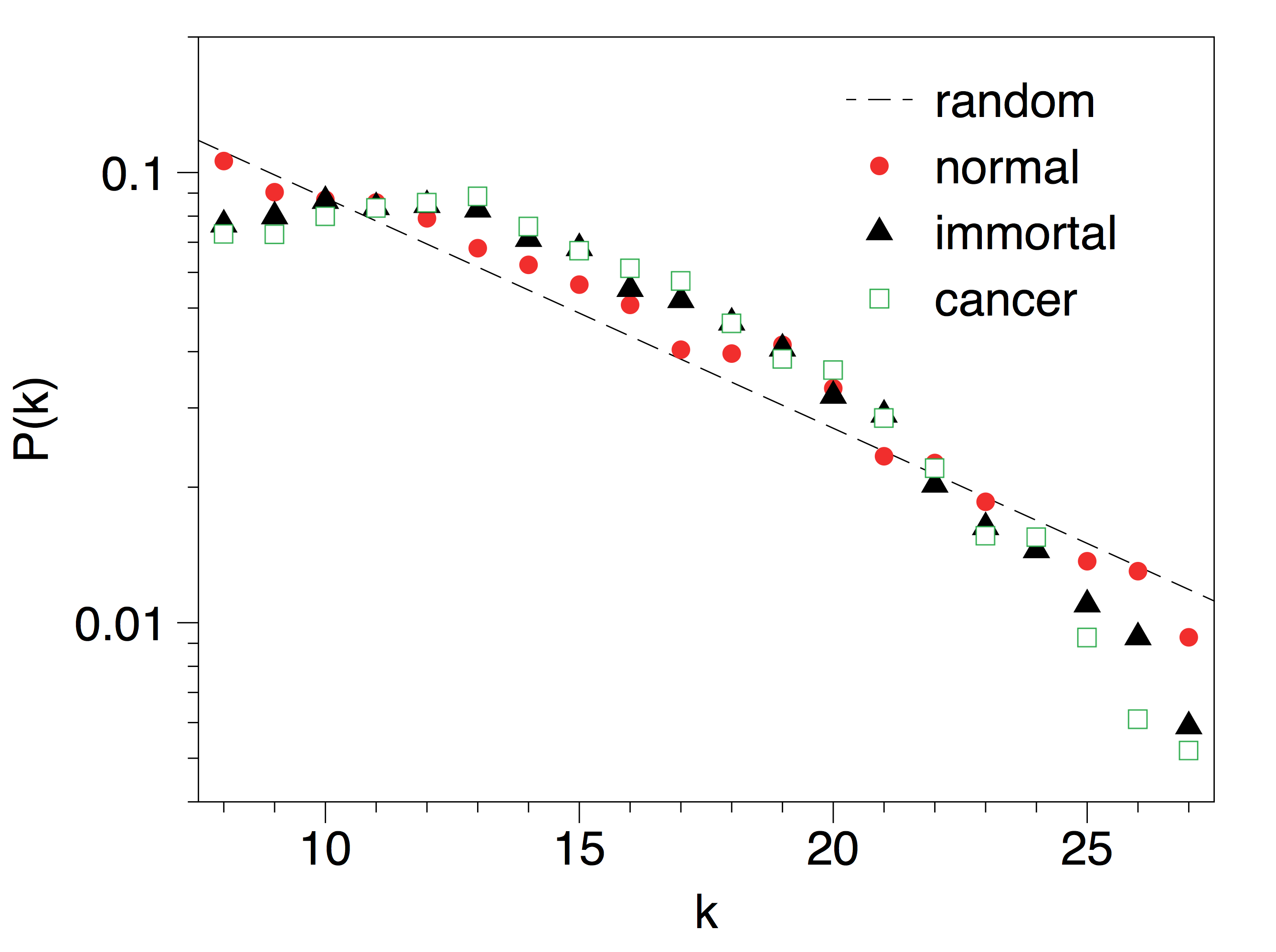}
\includegraphics[width=0.45\columnwidth]{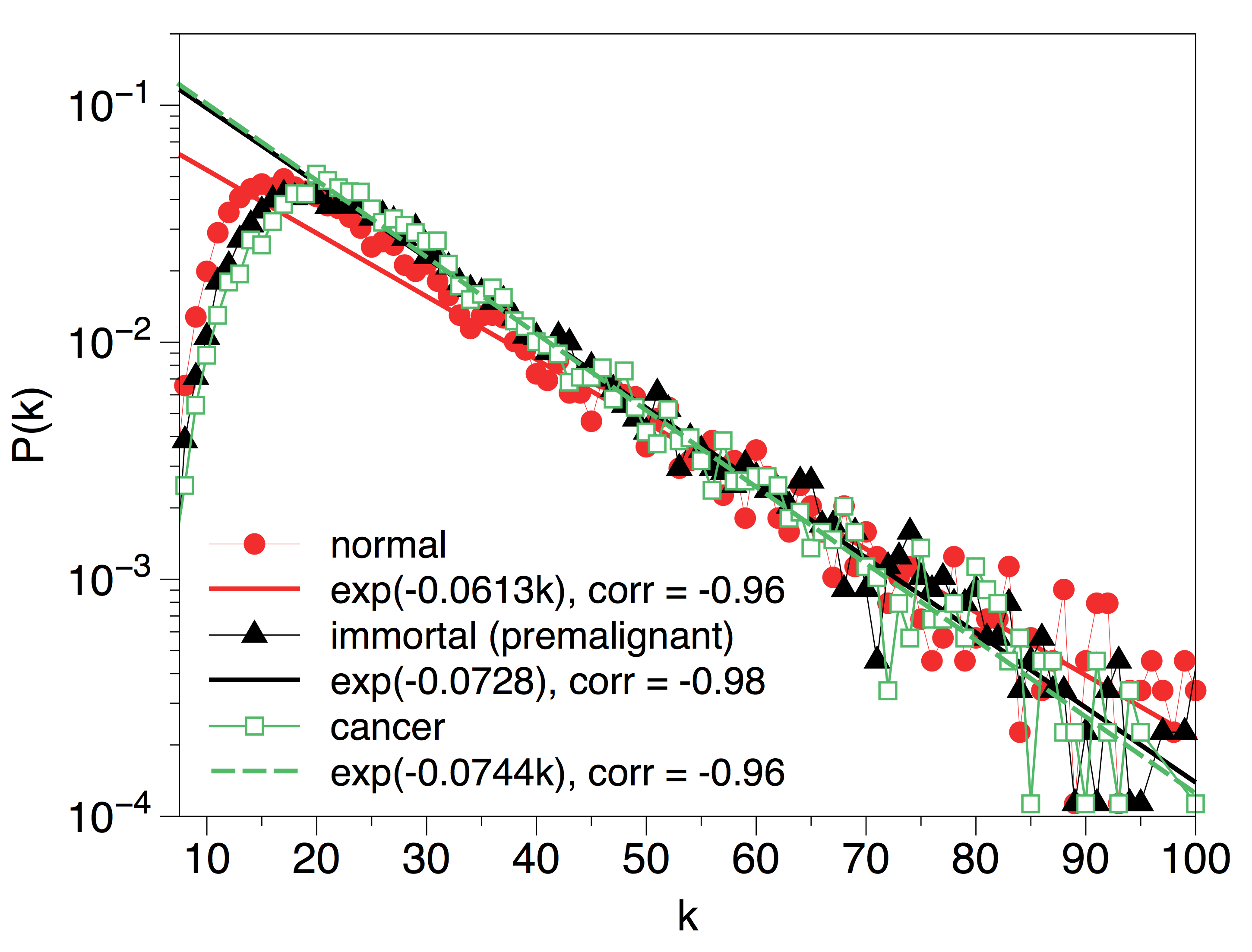}
\caption{{\it (Top panels)} Grayscale atomic  force microscopy images of normal (left), immortal (premalignant, middle) and cancer (malignant, right) cervical epithelial cells (extracted from \cite{cancer_paper} after permission from I. Sokolov).
 {\it(Bottom, Left panel)} Semi-log plot of the degree distribution of IHVG$_8$ associated to the three images: normal (red dots), immortal (black triangles) and cancerous (green hollow squares) cells \cite{cancer_paper}. Normal cells display a distribution closer to an uncorrelated random field (eq.\ref{theorem} for $n=8$), and this preliminirary evidence suggests that the transition normal $\to$ immortal $\to$ cancer is paralleled by a systematic deviation from the random field case for most of the degrees. {\it (Bottom, Right panel)} Same plot for the degree distribution of IVG$_8$, where we have fitted exponential functions $\sim \exp(-\lambda k)$ to the tails of the distributions, suggesting $\lambda_{\text{normal}}<\lambda_{\text{immortal}}<\lambda_{\text{cancer}}$.}
\label{cancer}
\end{figure}

\noindent For $d=2$ (either using the canonical or FCC extension classes, or the order-$n$ class), a plethora of applications emerge, here we only enumerate and discuss a few: (i) {\it Image Processing}: a (grayscale) image is just a discrete scalar field. Once we extract the visibility graphs of a given image, can we use the topological properties of this graph to build feature vectors which can feed automatic classifiers for several statistical learning tasks involving images \cite{image_processing}? Can we define the distance between two images using graph kernels \cite{kernel} on the associated visibility graphs? 
(ii) {\it Physics of Interfaces}: can we provide a topological characterization of fractal surface growth \cite{barabasi}? Can we -for instance- account for spatial self-similar structures much in the same way the Hurst exponent of fractional Brownian motion was estimated with visibility graphs \cite{EPL} (a preliminary analysis via row-column visibility graphs has partly addressed this issue recently \cite{rowcolumn}). Furthermore, can we apply this methodology in biologically-relevant problems and beyond, for instance to classify tumoral or calli surfaces? (iii) {\it Urban Planning}: can we automatically cluster cities by only resorting to combinatorial properties extracted from their visibility graphs? And can we link such emerging clusters with architectural, historical or cultural properties of cities? (iv) {\it Random Matrix theory}: Is there a visibility graph characterization of different random matrix ensembles?\\

\noindent To illustrate the potential applicability of the method to the case of tumor description, in the left panel of figure \ref{cancer} we plot the degree distribution of the IHVG$_8$ associated to three atomic force microscopy (AFM) images ($94\times 94$ after grayscale preprocessing) of normal, immortal (premalignant) and cancer (malignant) cervical epithelial cells \cite{cancer_paper}. This very preliminary evidence suggests that the carcinogenesis transition normal $\to$ premalignant $\to$ cancer is paralleled in graph space by a systematic deviation of the degree distribution from the i.i.d. case. In the right panel of the same figure we plot the degree distribution associated to IVG$_8$, whose tails have been fitted to exponential functions $\sim \exp(-\lambda k)$ finding $\lambda_{\text{normal}}<\lambda_{\text{immortal}}<\lambda_{\text{cancer}}$ \cite{cancer2}. These are of course very preliminary results given simply for illustration, and future research should confirm their accuracy and their potential use for carcinogenesis description and early detection.\\

\noindent The most exciting application for higher dimensions $d\geq 2$ is perhaps on describing the spatial structure of generic energy landscapes \cite{PEL1} $V:{\bf x}\in \mathbb{R}^d\to \mathbb{R}$, where $d$ is the number of degrees of freedom. Typically, these fields describe an energy function whose minimum is associated to the macroscopic behavior of many-body systems, and play a major role in physics and chemistry. The structure of these fields is however rather messy. As a matter of fact, in spin glasses and other disordered systems their macroscopic properties do not necessarily relate directly to a configuration of minimal energy as the system gets trapped in local, metastable minima of this energy surface: in this sense the spatial distribution and overall structure of these minima (stationary points) gives valuable information on the system dynamical evolution. These energy surfaces are also of great interest in chemistry (Kramer's reaction rate theory for the thermally activated escape from metastable
states) and high energy physics (e.g. local minima of supersymmetric energy landscape corresponds to the field theory vacuum). The formalism presented here would enable the description of such energetic landscapes, opening a thread of questions such as: Can we classify different types of field theories only using combinatorial criteria on their energy landscapes? What is the spatial distribution of stationary points of different canonical disordered systems in the light of this new method?\\

\noindent To conclude, we hopefully made the case that to encode spatially extended structures in a combinatorial fashion is an enterprise that opens exciting theoretical questions as well as applications. The approach presented here is promising and there exist several possible avenues for future research, and we hope that these methods spark interest in some of these communities accordingly.


\acknowledgments{We thank I. Sokolov for granting permission to reproduce the images on normal, immortal and cancer cells. LL acknowledges funding from EPSRC Fellowship EP/P01660X/1.}

\begin{figure}[h]
\centering
\includegraphics[width=0.28\columnwidth]{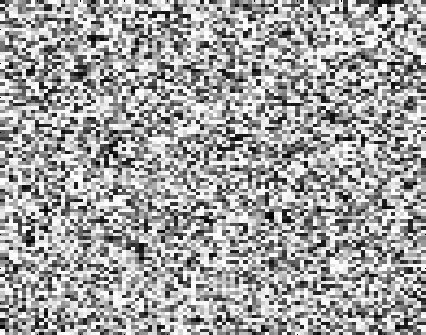}
\includegraphics[width=0.28\columnwidth]{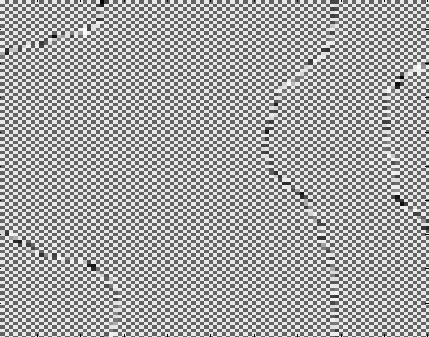}
\includegraphics[width=0.28\columnwidth]{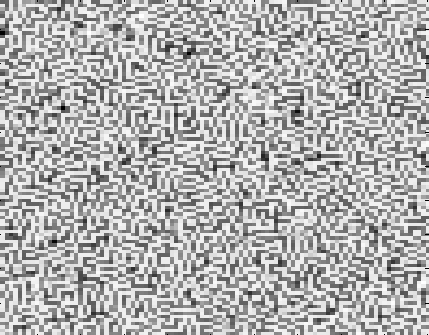}
\includegraphics[width=0.28\columnwidth]{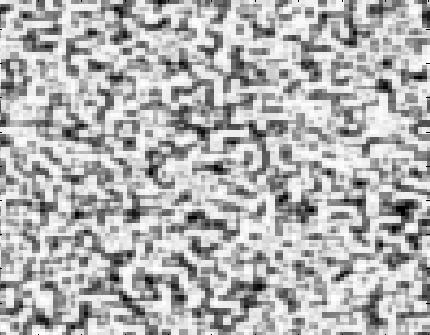}
\includegraphics[width=0.28\columnwidth]{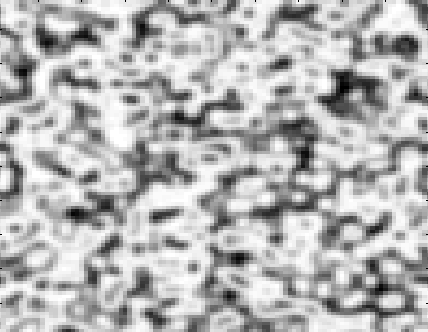}
\includegraphics[width=0.28\columnwidth]{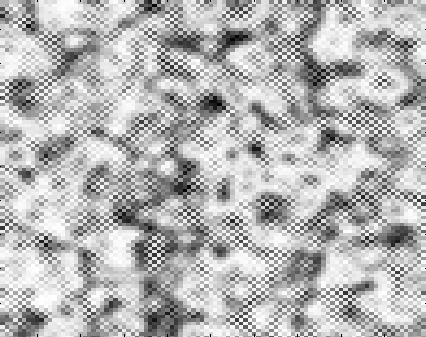}
\caption{Grayscale plots of 100x100 CMLs (eq.\ref{CMLeq}) for different values of $\epsilon$. From top left to bottom right, respectively: $\epsilon=0.05$ (Fully-developed turbulence), $\epsilon=0.15$ and $\epsilon=0.25$ (periodic structure), $\epsilon=0.4$ and $\epsilon=0.8$ (coherent structure), and $\epsilon=0.95$ (coexistence state with both coherent and periodic structures intertwined.}
\label{CML_epsilon}
\end{figure}



\bibliography{apssamp}

\begin{thebibliography}{10}
\bibitem{GT} B. Bollobas, {\it Modern Graph Theory} (Springer, 1998).
\bibitem{NS} M. Newman, {\it Networks: and introduction} (Oxford University Press, 2010).
\bibitem{Luque_Theorem} B. Luque, L. Lacasa, Canonical horizontal visibility graphs are uniquely determined by their degree sequence,
{\it Eur. Phys. J. B Sp. Top}. (in press).
\bibitem{libro_chaos} H. Kantz, T. Schreiber, {\it Nonlinear time series analysis} (Cambridge University Press).
\bibitem{PNAS} L. Lacasa, B. Luque, F.J. Ballesteros, J. Luque, and J.C. Nuno, From time series to complex networks: the visibility graph, {\it Proc. Natl. Acad. Sci. USA} {\bf 105}, 13 (2008).
\bibitem{image_processing} J. Iacovacci and L. Lacasa, Visibility graphs: a combinatorial framework for image processing (in preparation).
\bibitem{PRE} B. Luque, L. Lacasa, J. Luque, F.J. Ballesteros, Horizontal visibility graphs: exact results for random time series, {\it Phys. Rev. E} {\bf 80}, 046103 (2009).
\bibitem{severini} S. Severini, G. Gutin, T. Mansour, A characterization of horizontal visibility graphs and combinatorics on words, {\it Physica A} {\bf 390}, 12  (2011) 2421-2428.
\bibitem{flajo} P. Flajolet and M. Noy, Analytic combinatorics of non-crossing configurations, {\it Discrete Math.} {\bf 204} (1999) 203-229.
\bibitem{nonlinearity} L. Lacasa, On the degree distribution of horizontal visibility graphs associated to Markov processes and dynamical systems: diagrammatic and variational approaches, {\it Nonlinearity} {\bf 27}, 2063-2093 (2014).
\bibitem{nonstationary} L.Lacasa and R. Flanagan, Time reversibility from visibility graphs of non-stationary processes,
{\it Phys. Rev. E} {\bf 92}, 022817 (2015).
\bibitem{EPL} L. Lacasa, B. Luque, J. Luque and J.C. Nuno, The Visibility Graph: a new method for estimating the Hurst exponent of fractional Brownian motion, {\it EPL} {\bf 86}, 30001 (2009). 
\bibitem{rowcolumn} X Qin, P Xue, L Xin-Li, M Stephen, Y Hui-Jie, J Yan, W Jian-Yong, Z. Quin-Jung, Row-column visibility graph approach to two-dimensional landscapes, {\it Chinese Physics B} 23, 7 (2014).
\bibitem{original} A. Turner, M. Doxa, D. O'sullivan, and A. Penn, From isovists to visibility graphs: a methodology for the analysis of architectural space. {\it Environment and Planning B: Planning and design}, 28(1), 103-121 (2001).
\bibitem{CML} K. Kaneko, Overview of Coupled Map Lattices, {\it Chaos} {\bf 2}, 3(1992).
\bibitem{kernel} S.V. N. Vishwanathan, N.N. Schraudolph, R. Kondor and K.M. Borgwardt, Graph kernels, {\it Journal of Machine Learning Research} {\bf 11} (2010) pp.1201-1242.
\bibitem{barabasi}A.L. Barabasi and H.E. Stanley, {\it Fractal Concepts in Surface Growth} (Cambridge University Press, 1995).
\bibitem{PEL1} D. Wales, {\it Energy Landscapes : Applications to Clusters,
Biomolecules and Glasses} (Cambridge University Press, 2004).




\bibitem{jns} B. Luque, L. Lacasa, F. Ballesteros, A. Robledo,  Analytical properties of horizontal visibility graphs in the Feigenbaum scenario, {\it Chaos} {\bf 22}, 1 (2012) 013109.
\bibitem{quasi} B. Luque, A. N\'{u}\~{n}ez, F. Ballesteros, A. Robledo, Quasiperiodic Graphs: Structural Design, Scaling and Entropic Properties, {\it Journal of Nonlinear Science} {\bf 23}, 2, (2012) 335-342.
\bibitem{pre2013} A.M. N\'{u}\~{n}ez, B. Luque, L. Lacasa, J.P. G\'{o}mez, A. Robledo, Horizontal Visibility graphs generated by type-I intermittency, {\it Phys. Rev. E}, {\bf 87} (2013) 052801.

\bibitem{physics3} A. Aragoneses, L. Carpi, N. Tarasov, D.V. Churkin, M.C. Torrent, C. Masoller, and S.K. Turitsyn, Unveiling Temporal Correlations Characteristic of a Phase Transition in the Output Intensity of a Fiber Laser, {\it Phys. Rev. Lett.} 116, 033902 (2016).
\bibitem{fluiddyn0} M. Murugesana and R.I. Sujitha1, Combustion noise is scale-free: transition from scale-free to order at the onset of thermoacoustic instability, {\it J. Fluid Mech.} 772 (2015).
\bibitem{fluiddyn1}A. Charakopoulos, T.E. Karakasidis, P.N. Papanicolaou and A. Liakopoulos, The application of complex network time series analysis in turbulent heated jets, {\it Chaos} 24, 024408 (2014).
\bibitem{fluiddyn2} P. Manshour, M.R. Rahimi Tabar and J. Peinche, Fully developed turbulence in the view of horizontal visibility graphs, {\it J. Stat. Mech.} (2015) P08031.

\bibitem{physics2} RV Donner, JF Donges, Visibility graph analysis of geophysical time series: Potentials and possible pitfalls, {\it Acta Geophysica} 60, 3 (2012).





\bibitem{suyal} V. Suyal, A. Prasad, H.P. Singh, Visibility-Graph Analysis of the Solar Wind Velocity, \textit{Solar Physics} \textbf{289}, 379-389 (2014)
\bibitem{Zou} Y. Zou, R.V. Donner, N. Marwan, M. Small, and J. Kurths, Long-term changes in the north-south asymmetry of solar activity:
a nonlinear dynamics characterization using visibility graphs, \textit{Nonlin. Processes Geophys.} \textbf{21}, 1113-1126 (2014).

\bibitem{physio1} J.F. Donges, R.V. Donner and J. Kurths, Testing time series irreversibility using complex network methods, {\it EPL} 102, 10004 (2013).
\bibitem{meditation_VG} S. Jiang, C. Bian, X. Ning and Q.D.Y. Ma, Visibility graph analysis on heartbeat dynamics of meditation training, {\it Appl. Phys. Lett.} \textbf{102} 253702 (2013).


\bibitem{neuro} M Ahmadlou, H Adeli, A Adeli, New diagnostic EEG markers of the Alzheimer's disease using visibility graph, {\it J. of Neural Transm.} 117, 9 (2010).

\bibitem{ryan1} R. Flanagan and L. Lacasa, Irreversibility of financial time series: a graph-theoretical approach,
{\it Physics Letters A} 380, 1689-1697 (2016)
\bibitem{multivariate} L. Lacasa, V. Nicosia, V. Latora, Network Structure of Multivariate Time Series, {\it Sci. Rep.} \textbf{5}, 15508 (2015)


\bibitem{motifs} J. Iacovacci and L. Lacasa, Sequential visibility-graph motifs, {\it Phys. Rev. E} 93, 042309 (2016)

\bibitem{cancer_paper} M.E. Dokukin, N.V. Guz, C.D. Woodworth, and I. Sokolov, Emerging of fractal geometry on surface of human cervical epithelial cells during progression towards cancer, {\it New journal of physics} 17,3 (2015).
\bibitem{cancer2} M.E. Dokukin, N.V. Guz, R. M. Gaikwad, C.D. Woodworth, and I. Sokolov, Cell Surface as a Fractal: Normal and Cancerous Cervical Cells Demonstrate Different Fractal Behavior of Surface Adhesion Maps at the Nanoscale
{\it Phys. Rev. Lett.} 107, 028101 (2011).

\end{thebibliography}

\end{document}